\documentclass[aps,preprint,showpacs,preprintnumbers,amsmath,amssymb]{revtex4}
\usepackage{graphicx}
\usepackage{url} 
\begin{document}
\def\beq{\begin{equation}}
\def\eeq{\end{equation}}
\def\beqn{\begin{eqnarray}}
\def\eeqn{\end{eqnarray}}

\def\r {{\bf r}}
\def\n {{\bf n}}
\def\d {{\bf d}}
\def\C {{\bf C}}
\def\vC {{\bf C}}
\def\vphi  {\mbox{\boldmath $\phi$}}
\def\bcalH {\mbox{\boldmath $\mathcal H$}}

\def\brho {\mbox{\boldmath $\rho$}}

\def\bphi {\mbox{\boldmath $\phi$}}

\title{Accurate multi-boson long-time dynamics in triple-well periodic traps}
\author{Alexej I.\ Streltsov\footnote{E-mail: Alexej.Streltsov@pci.uni-heidelberg.de}}
\author{Kaspar Sakmann\footnote{E-mail: Kaspar.Sakmann@pci.uni-heidelberg.de}}
\author{Ofir E.\ Alon\footnote{E-mail: Ofir.Alon@pci.uni-heidelberg.de}} 
\author{Lorenz S.\ Cederbaum\footnote{E-mail: Lorenz.Cederbaum@pci.uni-heidelberg.de}} 
\affiliation{Theoretical Chemistry, Institute of Physical Chemistry,
Heidelberg University, Im Neuenheimer Feld 229, 69120 Heidelberg, Germany}
\date{\today}

\begin{abstract}
To solve the many-boson Schr\"odinger equation
we utilize the Multiconfigurational time-dependent Hartree method for bosons (MCTDHB).
To be able to attack larger systems and/or to propagate the solution for longer times,
we implement a parallel version of the MCTDHB method thereby realizing the recently proposed [Streltsov {\it et al.} arXiv:0910.2577v1]
novel idea how to construct efficiently the result of the action of the Hamiltonian on a bosonic state vector.
We study the real-space dynamics of repulsive bosonic systems made of $N=12$, 51 and 3003 bosons in triple-well periodic potentials.
The ground state of this system is three-fold fragmented.
By suddenly strongly distorting the trap potential, the system performs complex many-body quantum dynamics.
At long times it reveals a tendency to an oscillatory behavior around a threefold fragmented state.
These oscillations are strongly suppressed and damped by quantum depletions.
In spite of the richness of the observed dynamics, the three time-adaptive orbitals of MCTDHB($M=3$) are
capable to describe the many-boson quantum dynamics of the system for short and intermediate times.
For longer times, however, more self-consistent time-adaptive orbitals are needed to correctly describe 
the non-equilibrium many-body physics. The convergence of the MCTDHB($M$) method with the number $M$
of self-consistent time-dependent orbitals used is demonstrated.
\end{abstract}
\maketitle

\section{Introduction}\label{secI}

Quantum dynamics of bosonic systems \cite{Pit_Stri_book} is a highly active branch of modern physics.
It plays a crucial role in experiments with trapped ultra cold atomic or molecular clouds
especially when external trap potentials are varied in time. Theoretical predictions 
on the dynamics of quantum systems in these cases are highly desirable.
Quantum dynamics is usually governed by the time-dependent Schr\"odinger equation
\cite{Pit_Stri_book,Book_dynamics1,Book_dynamics2,Nuclear_book,Book_dynamics3,Book_dynamics4,Book_MCTDH,CI}.
To solve it means to define an initial condition, i.e., initial wave packet, and
to find the evolution of this initial state in time, i.e., to propagate it.
In this work we are interested in the many-body dynamics of
quantum systems made of indistinguishable bosons.
Specifically, to solve the time-dependend multi-boson Schr\"odinger equation
we use a the Multiconfigurational time-dependent Hartree
method for bosons (MCTDHB), developed recently in Refs.~\cite{MCTDHB1,MCTDHB2}.
The MCTDHB($M$) is a very efficient and successful propagation tool,
because it utilizes a very powerful ansatz for $N$-boson wave function,
defined in a complete Fock subspace, and spanned by all possible many-body basis functions (permanents) 
appearing as permutation of $N$ bosons over $M$ time-dependent self-consistent orbitals.

The key idea of the original MCTDH technique Ref.~\cite{MCTDH} was to consider as many-body basis
functions the Hartree products build of $N$ time-dependent orbitals, which can be, in principle, identical.
This method has been widely and successfully applied (see Ref.~\cite{Book_MCTDH}) to study multi-dimensional
quantum dynamics of polyatomic molecules, where each vibrational
mode has been assigned to a distinguishable quasi-particle.
Lately, it was realized that by applying anti-symmetrization operation to the Hartree products
one can study dynamics of fermionic systems, so the original MCTDH method
has been expanded in Ref.~\cite{MCTDHF} to the MCTDHF, here the letter ``F'' means fermions.
Analogously, we took symmetrized Hartree products (permanents) as many-body basis functions in Ref.~\cite{MCHB}
and showed their applicability for stationary multi-boson problems and, for the first time in this context,
reformulated the representation of the working equations in terms of elements of the reduced one- and two-body density matrices.
This reformulation has provided a deeper insight onto the underlying structure of the MCTDH ``ideology'' and allowed us to
derive in Refs.~\cite{MCTDHB1,MCTDHB2} closed-from MCTDHB equations for time-dependent bosonic problems and
extend the MCTDH ideas further in Refs.~\cite{Unified_paper,MIX} to mixed systems of bosons, fermions and spinors,
as well as to systems with particle conversion Ref.~\cite{MCTDH_CONV}.
In the latter case particles of one kind, e.g., atoms, are allowed to undergo transformation
and build up particles of another kind, i.e., molecules, and vice-versa.

The developed MCTDHB theory has been successfully applied to study 
a splitting of a repulsive bosonic cloud by ramping-up a potential barrier in Ref.~\cite{MCTDHB1}.
An optimal control of this process has been addressed in Ref.~\cite{MCTDHB_Joerg}.
The first- and second-order correlation functions of the split sub-clouds have been analyzed in Ref.~\cite{Corelation}.
In ultra-cold atomic clouds with attractive inter-boson interactions
the split phenomena take even more interesting forms.
In Ref.~\cite{Fragmenton} we have shown that by pumping energy into
an initially-coherent localized attractive bosonic cloud
one can fragment it into two dynamically stable sub-clouds, which are still strongly coupled.
Moreover, attractive bosonic systems also support another kind of dynamically stable
fragmented states -- the famous quantum superposition states.
These can be formed by scattering an attractive cloud from a potential barrier,
as was predicted independently by two groups 
using absolutely different many-body methods ~\cite{Caton_I_Castin,Caton_I}.
In Ref.~\cite{Caton_II} within the framework of the MCTDHB method we have shown
that there is a much more efficient way to generate the quantum superposition states 
-- instead of scattering an initially-coherent attractive cloud from the potential barrier
one can thread it by a barrier.
The MCTDH method is so numerically stable and powerfull that one can use the many-body basis sets made of
non-symmetrized Hartree products, originally proposed to describe distinguishable particles,
to study systems of indistinguishable particles.
The configurational spaces in these cases, however, are much larger than actually needed restricting
the applicability of such a scheme to few-particle problems.
Nevertheless, numerically-exact studies of few-boson problems using this scheme
have been successfully done in Refs.~\cite{Peter,Axel}.

To specify the many-body wave function in MCTDHB means to provide two sets of complex numbers --
a set of expansion coefficients, i.e., a vector-array of the size of the respective Fock subspace, and
a set of $M$ one-particle functions usually defined on a real space grid.
Both these sets are time-dependent, i.e., they are changing in time.
The time-evolution of the expansion coefficients in the Fock space
and the orbitals in space are governed by the respective MCTDHB equations \cite{MCTDHB1,MCTDHB2}.
In Ref.~\cite{Mapping_I} we have developed a novel, effective and general algorithm to operate
with state vectors in Fock space, without resorting
to matrix representation of operators and even to its elements.
Now our goal is to utilize, implement, and apply these ideas to the MCTDHB method.
In particular, we expect that a parallel implementation of the MCTDHB method
augmented by the proposed algorithm allows us to propagate the time-dependent Schr\"odinger
equation faster and more efficiently for larger configurational spaces, and thus be able to attack bigger systems,
and to propagate for longer times.

In this paper we apply the MCTDHB method augmented by the invented algorithm ~\cite{Mapping_I}
to study real-space dynamics of repulsive bosonic systems made of $N=12$, 51 and 3003 bosons
in triple-well traps with periodic boundaries.
As initial wave packets we take fully threefold fragmented states
and displace them suddenly and strongly out of equilibrium.
The periodic trap used for the propagations supports at least six bound eigenstates per well,
i.e., six bands are available for the quantum dynamics.
As a consequence of the displacement applied, the energies per particle of the
evolving wave packets are located between the fourth and fifth bands which implies strongly non-equilibrium dynamics.
On the one hand, we would like to see how the properties of these highly non-equilibrium systems change in time and
what are the most relevant physical phenomena behind.
For example, to clarify the role of finite $N$ effects we analyze and compare
the quantum dynamics of the systems made of different number of bosons
but characterized by very similar interparticle interaction energies.
On the other hand, we also plan to demonstrate the convergence of the MCTDHB($M$) method,
for short, intermediate and long propagation times.
More specifically, we compare the computations done
at different levels $M$ of the MCTDHB($M$) theory, i.e., by
systematically increasing the number $M$ of time-dependent self-consistent orbitals.
The key methodological consequence of this study is to show that due to the algorithm used
one can tackle large configurational spaces thus opening many new opportunities for the MCTDHB method.

The structure of the paper is as follows.
In section \ref{secII} we consider a general many-boson system in real space, define its Hamiltonian and
the $N$-boson state vector (wave function).
The basics of the multiconfigurational time-dependent Hartree method for bosons (MCTDHB) are given in subsection \ref{secII.1}.
The implementation of the novel idea how to operate by a bosonic Hamiltonian on a state vector
within the MCTDHB method is discussed in subsection \ref{secII.2}.
In particular, we describe how to map and address permanents (configurations),
how to operate with one- and two-body operators as well as with total Hamiltonian, and
how to construct the respective expectation values.
Application of the MCTDHB method to many-boson quantum dynamics in triple-well traps is illustrated in section \ref{secIII}.
In subsection \ref{secIII.1} we provide the specification of the systems under consideration --
initial and final trap potentials, initial wave packets, interparticle interactions, etc. 
The detailed analysis of the evolving wave packets for the systems made of $N=12$, 51 and 3003 bosons are reported in
the subsections \ref{secIII.2}, \ref{secIII.3}, and \ref{secIII.4} respectively.
Section \ref{secIV} summarizes and concludes.

\section{Theory}\label{secII}

The quantum dynamics of a system made of $N$ bosons is governed by the many-body time-dependent Schr\"odinger equation.
This equation describes the evolution of a many-boson wave function $\Psi(t)$ in time:
\beqn
 \hat H \Psi(t) & = & i \frac{\partial \Psi(t)}{\partial t}, \qquad  \label{MBSE} \\ 
 \hat H(\r_1,\r_2,\ldots,\r_N) &= & \sum_{j=1}^{N} \hat h(\r_j) +  \sum_{k>j=1}^N \hat W(\r_j-\r_k). \label{MBSE1}
\eeqn
were $\hbar=1$, $\r_j$ is the coordinate of the $j$-th boson, $\hat h(\r) = \hat T(\r) + \hat V(\r)$ is the one-body Hamiltonian
containing kinetic and potential energy terms, and $\hat W(\r_j-\r_k)$ describes the pairwise interaction between the $j$-th and $k$-th bosons.
In the most general case, the one-body potential $\hat V(\r)$ and the two-body interaction $\hat W(\r_j-\r_k)$ and, hence,
the many-boson Hamiltonian $\hat H$ itself may be time-dependent.
To solve Eq.~(\ref{MBSE}) means to specify initial conditions at $t=t_0$ and propagate the ansatz in time.
More precisely, we have to specify the many-body wave function $\Psi(t=t_0)$, i.e., an object which depends on coordinates $\r_i$ of
each of $N$ particles and by using Eq.~(\ref{MBSE}) find the many-boson wave function at desired time $\Psi(t)$.

\subsection{Multiconfigurational time-dependent Hartree method for bosons (MCTDHB)}\label{secII.1}

General analytical solutions of the time (in)dependent Schr\"odinger are unknown, therefore,
different numerical approaches are in use \cite{Pit_Stri_book,Book_dynamics1,Book_dynamics2,Nuclear_book,Book_dynamics3,Book_dynamics4,Book_MCTDH}.
In this section, we consider one of them -- the multiconfigurational time-dependent Hartree method for bosons (MCTDHB).
The general formulation and derivation of the MCTDHB method have been given elsewhere, see Refs.~\cite{MCTDHB1,MCTDHB2} for the details.
Here, we present the final equations and give some remarks on the numerical realization
of the novel ideas in \cite{Mapping_I} how to operate in a configurational (Fock) space without resorting to the
matrix representation of the respective operators.

The MCTDHB ansatz, i.e., the proposed structure of the many-boson wave function
is a linear combination of time-dependent many-body basis functions $|\vec{n};t\rangle$:
\beq\label{MCTDHB_Ansatz}
\left|\Psi(t)\right> =
\sum^{N_{\mathit{conf}}}_{\vec{n}}C_{\vec{n}}(t)\left|\vec{n};t\right>.
\eeq
For a system of indistinguishable bosons we take permanents as many-body basis functions.
Every permanent is constructed as a symmetrized Hartree product of $N$ one-particle functions (orbitals):
\beq\label{basic_permanents_x}
\langle \r_1,\r_2,\r_3,\cdots,\r_N |\vec{n};t\rangle =\hat{\cal S}\phi_1(\r_1,t)\cdots\phi_1(\r_{n_1},t)\phi_2(\r_{n_1+1},t)
\cdots\phi_M(\r_{n_1+n_2+\cdots+n_M},t),
\eeq
where $n_1$ bosons reside in the first orbital, $n_2$ bosons reside in second one, etc.,
and $n_M$ bosons reside in the $M$-th orbital. The possible occupation patterns $\vec{n}$,
also called configurations, are constrainted only by the condition
that the total number of particles is conserved $n_1+n_2+\cdots+n_M=N$.
So, the complete configurational space of the standard MCTDHB($M$) method
is formed by all possible permanents obtained by distributing $N$ bosons over $M$ orbitals \cite{CC_paper}:
\beq\label{Nconf}
N_{\mathit{conf}}=\binom{N+M-1}{N},
\eeq
where $\binom{n}{k}=\frac{n!}{k!(n-k)!}$.
The specification MCTDHB($M$) means that $M$ self-consistent time-adaptive orbitals have been used
in the computation and the size of the respective Fock space, i.e., the total number of the configurations
is determined by Eq.~(\ref{Nconf}).

At any given time the many-boson MCTDHB($M$) wave function $\left|\Psi(t)\right>$ is known
if the set of expansion coefficients $\{C_{\vec{n}}(t)\}\equiv\vC(t)$, i.e., a complex array of $N_{\mathit{conf}}$ elements and
the set of $M$ one-particle functions $\{\phi_i(\r,t)\}$ 
are specified.
Here, we do not discuss how one-particle functions are represented because this depends on
the dimensionality of the problem and on the particular realization of the code.
For example, in 1D cases, one can represent each of these functions on a $one$-dimensional grid as
a vector-array of complex numbers.

The MCTDHB($M$) ansatz Eq.~(\ref{MCTDHB_Ansatz}) possesses a standard feature of any many-body time-dependent expansion --
its expansion coefficients $\vC(t)$ depend on time.
It reflects the basic idea of quantum propagation: The many-body wave packet is composed of
contributions of very many states (configurations) and these contributions are changing in time.
The ansatz Eq.~(\ref{MCTDHB_Ansatz}) has an additional, the MCTDHB- specific feature -- dependency of the orbitals,
i.e., one-particle functions $\{ \phi_i(\r,t) \}$ used to construct the permanents Eq.~(\ref{basic_permanents_x}) on time.
The orbitals $\{ \phi_i(\r,t) \}$ of the evolving many-boson wave packet
are optimized at every point in time according to the Dirac-Frenkel variational principle \cite{DF1,DF2},
changing, thereby, the many-body basis permanents.
We shall see later how this additional MCTDHB-specific feature allows one to span effectively
much larger configurational spaces in comparison to methods using fixed, i.e., time-independent orbital basis sets.

In the MCTDHB theory the expansion coefficients $\vC(t)$ and the shapes of one-particle functions $\{\phi_i(\r,t)\}$
are considered as independent variational parameters, see Refs.~\cite{MCTDHB1,MCTDHB2}.
We substitute the MCTDHB($M$) ansatz Eq.~(\ref{MCTDHB_Ansatz}) into the Hamiltonian
and introduce Lagrange multipliers to guarantee the normalization
of the total wave function as well as orthonormalization of the orbitals.
Then, by utilizing the Dirac-Frenkel time-dependent variational principle \cite{DF1,DF2}
we get a coupled system of integro-differential equations 
which have to be solved (integrated) simultaneously. According to the derivation,
the coupled system is composed of two sets of the variational equations,
one for the expansion coefficients $\vC(t)$ and another one for the one-particle functions $\{ \phi_i(\r,t) \}$.
In numerical implementations of these equations, it is usually assumed that
during the elementary propagation of the first set of variational parameters, say expansion coefficients from $\vC(t)$ to $\vC(t+\Delta t)$,
the second one, i.e., the orbital set $\{ \phi_i(\r,t) \}$, is kept fixed,
and vice-versa, during the propagation of the orbitals from
$\{ \phi_i(\r,t) \}$ to $\{ \phi_i(\r,t+\Delta t) \}$ the $\vC(t)$-set is kept fixed.
Of course, the $\Delta t$ has to be sufficiently small.

The first set of the MCTDHB equations describes the evolution of $N_{\mathit{conf}}$ expansion coefficients
and can be rewritten as:
\beq\label{MCTDHB_CI}
i\frac{\partial \vC(t)}{\partial t}= \bcalH \vC(t),
\eeq
where $\bcalH =\left\{\left<\n;t\left|\hat H\right|\n';t\right>\right\}$
is the matrix representation of the Hamiltonian.
In this equation it is assumed that at each point in time,
the shapes of the orbitals $\{ \phi_i(\r,t) \}$ are known and, therefore, 
during an elementary numerical propagation time step they remain unchanged, i.e., fixed.
Hence, the original real-space Hamiltonian operator [see Eq.~(\ref{MBSE1})] is represented now in the second quantized form:
\beq\label{MB_Ham}
  \hat H =
 \sum_{k,q} h_{kq}  b_k^\dag b_q
 + \frac{1}{2}\sum_{k,s,q,l} W_{ksql} b_k^\dag b_s^\dag b_l b_q,
\eeq
where the elements $h_{kq}$ of the one- and $W_{ksql}$ of the two-body operators are
directly available as integrals over the discussed above orbitals:
\beqn\label{matrix_elements}
 h_{kq} &=& \int \phi_k^\ast(\r,t) \hat h(\r) \phi_q(\r,t) d\r, \nonumber \\
W_{ksql} &=& \int \!\! \int \phi_k^\ast(\r,t) \phi_s^\ast(\r',t) \hat W(\r-\r')
 \phi_q(\r,t) \phi_l(\r',t) d\r d\r'.
\eeqn

The second set of the MCTDHB equations, also called equations-of-motion for orbitals,
describes the evolution of $M$ orbitals $\phi_j(\r,t)$ and can be written as:
\beqn\label{MCTDHB_PSI}
  i\left|\dot\phi_j\right> & = & \hat {\mathbf P} \left[\hat h \left|\phi_j\right>  + \sum^M_{k,s,l,q=1}
  \left\{\brho(t)\right\}^{-1}_{jk} \rho_{kslq} \hat{W}_{sl} \left|\phi_q\right> \right], \, j=1,\cdots,M,  \\
\qquad \hat {\mathbf P} & = & 1-\sum_{j'=1}^{M}\left|\phi_{j'}\left>\right<\phi_{j'}\right|, \nonumber \\
\hat W_{sl} & = & \int \phi_s^\ast(\r',t) \hat W(\r-\r')
\phi_l(\r',t) d\r', \nonumber 
\eeqn
where the expansion coefficients $\vC$ are assumed to be known and
remain fixed (i.e., unchanged) during an elementary numerical propagation time step.
There are no individual contributions of the expansion coefficients
to these equations-of-motion, they enter only as integrals \cite{MCTDHB1,MCTDHB2} 
via the elements $\rho_{kq}$ and $\rho_{ksql}$ of the reduced one- and two-body density matrices:
\beqn\label{MCTDHB_DNS}
 \rho_{kq}&=&\left<\Psi\right| b_k^\dag b_q  \left|\Psi\right>,  \\
 \rho_{kslq}&=&\left<\Psi\right| b_k^\dag b_s^\dag b_l b_q  \left|\Psi\right>. \nonumber
\eeqn
Since the orbitals $\{ \phi_i(\r;t) \}$ are time-dependent functions,
the integrals $h_{ks}$ and $W_{ksql}$, computed according to Eq.~(\ref{matrix_elements}) become 
time-dependent functions even for time independent Hamiltonians.
Hence, the price of using the self-consistent time-adaptive orbitals is that at every elementary numerical propagation time step
one has to update these integrals, i.e., the second quantized form of the Hamiltonian, Eq.~(\ref{MB_Ham}).

Having the MCTDHB($M$) wave function $\left|\Psi(t)\right>$ at every given time,
i.e., the $N_{\mathit{conf}}$ complex numbers $\vC(t)$ and
$M$ one-particle functions $\{ \phi_i(\r,t) \}$,
we have an access to the elements of the reduced one-, two- or higher-order density matrices and
can construct, thereby, respective correlation functions \cite{Corelation} or compute any desired properties.
Specifically, in this work we construct and diagonalize the reduced one-body density matrix:
\beqn\label{NO}
\rho(\r,\r';t)&=& \left<\Psi\right| \hat \Psi^\dag(\r') \Psi(\r) \left|\Psi\right>= \nonumber \\
              &=& \sum_{k,q} \rho_{kq}  \phi_k^\ast(\r,t) \phi_q(\r',t)=N\sum_{k} n_k(t) \phi_k^{\ast NO}(\r,t) \phi_k^{NO}(\r',t),
\eeqn
the obtained eigenvalues $n_k(t)$ and eigenvectors $\phi_k^{NO}(\r,t)$ are called
natural occupation numbers and natural occupation orbitals, respectively.
Hence, this diagonalization procedure is often called natural analysis.
A simple visualization of the reduced one-body density is obtained by plotting 
only its diagonal part $\rho_{\mathit{MB}}(\r)\equiv\rho(\r,\r;t)$, also referred to as density.

Summarizing, there are $N_{\mathit{conf}}$ MCTDHB equations (\ref{MCTDHB_CI}) describing the evolution
of the expansion coefficients, and $M$ non-linear integro-differential
equations (\ref{MCTDHB_PSI}) describing the evolution of the one-particle functions.
To propagate the initial state, we divide time onto small elementary slices or steps.
During every elementary time step either set of expansion coefficients
or one-particle functions is assumed to be unchanged. This allows within a given time step
to integrate the MCTDHB Eqs.~(\ref{MCTDHB_CI}) and Eqs.~(\ref{MCTDHB_PSI}) independently,
see Refs.~\cite{MCTDHB1,MCTDHB2} for more details.
The evolution of the one-particle functions, i.e., propagation of the respective Eqs.~(\ref{MCTDHB_PSI}) can be done by using
general variable-order integrators, such as the Adams-Bashforth-Moulton (ABM) predictor-corrector integrator \cite{ABM}.
To propagate Eqs.~(\ref{MCTDHB_CI}) for expansion coefficients
we can use the short iterative Lanczos (SIL) scheme \cite{SIL}, because
in the present work we deal with hermitian Hamiltonians, i.e., with real traps and real interparticle interaction potentials.
The key ingredient of the SIL technique is a $K$-dimensional Krylov subspace, formed by repeated operations of the Hamiltonian:
$\left\{ |\Psi \rangle, \hat H^1 | \Psi \rangle, \hat H (\hat H | \Psi \rangle),\cdots,\hat H^{K-1} | \Psi \rangle \right\}$.
To integrate the configurational part of the MCTDHB($M$) equations [see Eqs.~(\ref{MCTDHB_CI})] at a given point in time
we need these vectors only.

In a standard computational scheme to construct Krylov vectors
at every integration step one first recalculates the Hamiltonian matrix elements $\left<\n;t\left|\hat H\right|\n';t\right>$ 
which is a quite tedious task that even for a moderate-sized systems. And then by
multiplying this matrix with the complex vector-array of the expansion coefficients repeatedly
one builds up the Krylov subspace of the desired size.
In Ref.\cite{Mapping_I} we invented an alternative technique to compute
the action of the Hamiltonian on a state vector.
Now, we discuss how to implement the proposed ideas to the MCTDHB method
to obtain quickly and efficiently all the required vectors without (re)construction of
the respective Hamiltonian matrix at every point in time.
This allows, as we shall demonstrate below, for faster operations in larger configurational spaces.

\subsection{Details of implementation of MCTDHB}\label{secII.2}

The propagation of the MCTDHB wave function within a given time interval consists of two steps:
propagation of the expansion coefficients keeping the orbitals fixed, Eqs.~(\ref{MCTDHB_CI}),
and propagation of the orbitals keeping the expansion coefficients unchanged, see Eqs.~(\ref{MCTDHB_PSI}).
The key operation at the first step is the action of the Hamiltonian Eq.~(\ref{MB_Ham}) on a state vector $\left|\Psi\right>$,
having at hands the integrals $h_{kq}$ and $W_{ksql}$.
At the second step we need to know the elements $\rho_{kq}$ and $\rho_{ksql}$ of the reduced one- and two-body density matrices.
This subsection shows how to get these quantities by implementing and utilizing the ideas reported in Ref.~\cite{Mapping_I}.

As we have seen above, the set of the one-particle functions $\{ \phi_i(\r,t) \}$ 
and the set of expansion coefficients $\vC(t)$, i.e., the complex array of $N_{\mathit{conf}}$ elements,
specify the MCTDHB wave function $\left|\Psi(t)\right>$ at any given time.
We first introduce a simple and compact scheme to enumerate elements of the complex array $\vC$.
More strictly, we utilize the explicit scheme of Ref.~\cite{Mapping_I} to map $M$ integers $n_1,n_2,n_3,\ldots,n_M$
characterizing each configuration to one integer $J$ addressing it as a coordinate of the state vector.
The address $J$ of the bosonic configuration $|n_1,n_2,n_3,\cdots,n_M \rangle$
in the state vector is computed as:
\begin{equation} \label{j-numbering}
J(n_1,n_2,\cdots,n_{M-1},n_M)=1+\sum_{k=1}^{M-1} \binom{N+M-1-k-\sum_{l=1}^{k}n_l}{M-k}.
\end{equation}
Technically, it means that for any given set of $M$ integer occupation numbers $\vec{n}\equiv(n_1,n_2,n_3,\cdots,n_M)$
we can immediately compute one integer -- its address $J$.
Consequently, using this mapping we can access any desired element of the complex array $\vC$
by providing the respective set of the occupation numbers.

The next question is how to access the elements of the $\vC$ array.
Let us take the definition of the MCTDHB($M$) ansatz (\ref{MCTDHB_Ansatz}) and rewrite it:
\beq\label{J_Psi}
\left|\Psi\right> =
\sum^{N_{\mathit{conf}}}_{J=1}C_{J}\left|J(\n)\right>,
\eeq
where the summation index $J$ running over all $N_{\mathit{conf}}$ configurations
is associated with the above defined mapping index Eq.~(\ref{j-numbering}).
To access every $C_{J}$ element by starting one {\it DO-}loop over $J$ is inconvenient, because
one has to, first, restore $M$ occupation numbers $(n_1,n_2,n_3,\cdots,n_M)$ corresponding to every given address $J$
at every step, i.e., to solve the respective inverse problem. That is in principle possible but not necessary.
To avoid it, we start, instead, $M$ nested {\it DO-}loops running over occupation numbers $n_i$
with the condition that $n_1+n_2+n_3+\cdots+n_M=N$, the index $J$ is immediately available by applying Eq.~(\ref{j-numbering}).
Hence, in the case of an $M$-loop implementation there is no need to solve the inverse problem at all.
From now on, by writting a sum over index $J$ we actually mean
that in practical implementations $M$ nested loops over $n_i$ are running instead.
The repeated recomputations of the binomial coefficients needed by Eq.~(\ref{j-numbering}) can be avoided by
introducing a small static two-dimensional array and filling it only once by the respective binomial integers.
Hence, to find the address $J$ corresponding to a given set $(n_1,n_2,n_3,\cdots,n_M)$ of the 
occupation numbers one needs to sum up $M-1$ elements of this integer array.

Now, we approach the key issue -- how to find the result of the action
of the Hamiltonian Eq.~(\ref{MB_Ham}) on a bosonic state vector.
Within a standard quantum mechanical text-book approach, e.g., Ref.~\cite{CI},
to get $\hat H | \Psi \rangle$ one needs, first, to construct the respective Hamiltonian matrix
and then to multiply it by the complex vector of expansion coefficients $\vC$.
In Ref.~\cite{Mapping_I} we propose a simple alternative to it.
The Hamiltonian is a sum of one-body $\hat h =\sum_{k,q} h_{kq} b^\dag_kb_q$ 
and two-body $\hat W =\frac{1}{2}\sum_{k,s,q,l} W_{ksql} b^\dag_kb^\dag_sb_lb_q$ terms.
If the result of action of every basic one-body $b^\dag_kb_q$ and 
two-body $b^\dag_kb^\dag_sb_lb_q$ operator on a state vector is known,
the total action of the Hamiltonian is obtained by summing them up.
In ~\cite{Mapping_I} we discussed these operations in great details,
here we give only the final formulations.

\subsubsection{Actions of the basic one- and two-body operators}\label{secII.2.1}

In Ref.~\cite{Mapping_I} we have seen that the action of the $b^\dag_kb_q$ combination
of creation and annihilation operators on every configuration 
$\left|n_1,\cdots,n_k  ,\cdots,n_q  ,\cdots,n_M \right>\equiv\left|J\right>$
of the incoming state vector can be interpreted as a re-addressing of this configuration to another one $\left|n_1,\cdots,n_k+1,\cdots,n_q-1,\cdots,n_M \right>$ scaled by a statistics-dependent weight.
Therefore, using the introduced above mapping we can get the result of the action
of the $b^\dag_kb_q$ on every component of the bosonic state vector $\left|\Psi\right>$, for a given pair $k$ and $q$:
\beqn\label{one-body_oper}
\left| \Psi^{kq} \right> \equiv b^\dag_kb_q \left| \Psi \right> =
\sum^{N_{\mathit{conf}}}_{J}C^{kq}_{J}\left|J(\n)\right>, \\
C^{kq}_{J}=C_{J^{kq}}\sqrt{n_k}\sqrt{n_q+1} \nonumber.
\eeqn
As we have discussed above, instead of running one {\it DO-}loop over index $J$, we start here 
$M$ nested {\it DO-}loops over $n_i$. The occupation pattern
of the current configuration $J$ is explicitly known $(n_1,\cdots,n_k,\cdots,n_q,\cdots,n_M)$.
The occupation pattern of the configuration that we reffer to as $J^{kq}$
is obtained from the current one by reducing the occupation number of the
$k$-th orbital by one, i.e., now it becomes $n_k-1$ and by increasing the occupation number of the $q$-th orbitals by one $n_q+1$:
$(n_1,\cdots,n_k-1,\cdots,n_q+1,\cdots,n_M)$.
The change between $q$ and $k$ indices appears due to the change of the summation index, see Ref.~\cite{Mapping_I} for more details.
Using Eq.~(\ref{j-numbering}) we compute the addresses of the
current $J =J(n_1,\cdots,n_k,\cdots,n_q,\cdots,n_M)$
and referred $J^{kq}=J(n_1,\cdots,n_k-1,\cdots,n_q+1,\cdots,n_M)$ configurations.
The element $C^{kq}_{J}$ with address $J$ of the resulting state vector is obtained as a product of
the $C_{J^{kq}}$ component of the incoming state vector, having address $J^{kq}$, and the $\sqrt{n_k}\sqrt{n_q+1}$ prefactor.
This simple and straightforward methodology is ideally suitable for programming.

Moreover, within the framework of this method we also get a direct access to the elements $\rho_{kq}$ of the reduced one-body density matrix.
Indeed, according to the standard definition,
for a given state vector $\left|\Psi\right>$,
the one-body density matrix element reads:
\beqn\label{rho_ij}
\rho_{kq}= \left<\Psi\right| b^\dag_kb_q\left|\Psi\right>\equiv\left<\Psi\right| \left[ b^\dag_kb_q\left|\Psi\right> \right]
=\left<\Psi|\Psi^{kq}\right>=\sum^{N_{\mathit{conf}}}_{J}C^*_{J}C^{kq}_{J}.
\eeqn
Thus, the matrix elements $\rho_{kq}$ can be immediately computed as a dot-product
of the incoming $\left|\Psi\right>$ and resulting $\left|\Psi^{kq}\right>$ state vectors.

Similarly, we get the action of the general ($k\neq s \neq q \ne l$) two-body $b^\dag_kb^\dag_sb_lb_q$ operator on
the incoming state vector $\left|\Psi\right>$:
\beqn\label{two-body_oper}
\left|\Psi^{kslq}\right> \equiv b^\dag_kb^\dag_sb_lb_q\left|\Psi\right> =
\sum^{N_{\mathit{conf}}}_{J}C^{kslq}_{J}\left|J(\n)\right>, \\
C^{kslq}_{J}=C_{J^{kslq}}\sqrt{n_k} \sqrt{n_s}\sqrt{n_l+1} \sqrt{n_q+1} \nonumber ,
\eeqn
where the current index $J =J(n_1,\cdots,n_k,\cdots,n_s,\cdots,n_l,\cdots,n_q,\cdots,n_M)$ of the resulting vector
as well as the address of the incoming vector, which we refer to $J^{kslq}=J(n_1,\cdots,n_k-1,\cdots,n_s-1,\cdots,n_q+1,\cdots,n_l+1,\cdots,n_M)$ 
are computed using Eq.~(\ref{j-numbering}).
The respective element of the two-body density is obtained as a dot-product of the incoming $\left|\Psi\right>$ and 
resulting $\left|\Psi^{kslq}\right>$ state vectors:
\beqn\label{rho_ijkl}
\rho_{kslq}= \left<\Psi\right| b^\dag_kb^\dag_sb_lb_q \left|\Psi\right>\equiv\left<\Psi\right| \left[ b^\dag_kb^\dag_sb_lb_q\left|\Psi\right> \right]
=\left<\Psi|\Psi^{kslq}\right>=\sum^{N_{\mathit{conf}}}_{J}C^*_{J}C^{kslq}_{J}.
\eeqn
Obviously, in the discussed scheme the elements of the one- and two-body density matrices, needed by
the second part of the MCTDHB($M$) equations (\ref{MCTDHB_PSI}) are naturally available.
Clearly, the elements of three- and higher-body density matrices can be obtained in a very similar way.

The following strategy is implemented in the MCTDHB($M$) code to obtain
the result of the action of the Hamiltonian on a state vector. 
First, for a given vector $\left|\Psi\right>$, i.e., a given set of the expansion coefficients $\vC$,
the actions of every one- and two-body basic operators are evaluated using Eqs.~(\ref{one-body_oper}) or (\ref{two-body_oper}).
Each of these actions can be computed independently, implying effective parallelization.
Namely, the $\vC$ array, corresponding to the incoming vector is copied over all the available nodes.
Then, on every node, independently, we start $M$ nested loops
to get the resulting $\left|\Psi^{kq}\right>$ and $\left|\Psi^{kslq}\right>$ vectors,
corresponding to every pair $\{k,q\}$ and quartet $\{k,s,q,l\}$.
Second, by taking the dot-product of the incoming and the resulting state vectors, i.e., by using Eqs.~(\ref{rho_ij}) or (\ref{rho_ijkl})
we obtain the respective elements $\rho_{kq}$ and $\rho_{kslq}$ of the reduced one- and two-body density matrices.
Third, by multiplying all the elements of the resulting vectors by the corresponding integrals
we get the desired actions of the respective $h_{kq}\left|\Psi^{kq}\right>$ and $W_{ksql}\left|\Psi^{kslq}\right>$ terms
of the Hamiltonian on the incoming state vector.
Fourth, by summing up the resulting vectors from all the terms we obtain the total action of the Hamiltonian on the incoming state vector.
Finally, we easily compute the expectation value of the Hamiltonian $\langle \Psi| \hat H | \Psi \rangle$ as a dot-product
of the total resulting and incoming state vectors.

\subsubsection{Action of the Hamiltonian}\label{secII.2.2}

In principle, the desired vector $\hat H | \Psi \rangle$ can be obtained
by collecting actions from one- and two- body operators into two separate groups and summing only two resulting vectors afterwards.
Using the results of the previous subsections we determine the total action of all the one-body terms
on an initial state vector $\left|\Psi\right>$ as:
\beqn\label{H-psi_one-body}
\hat h \left|\Psi\right> &=&\sum_{k,q} h_{kq} \left [  b^\dag_kb_q \left|\Psi\right> \right ]
= \sum_{k,q} h_{kq}  \left|\Psi^{kq}\right>=\sum^{N_{\mathit{conf}}}_{J}C^{\hat h}_{J}\left|J(\n)\right>,  \nonumber \\
C^{\hat h}_J&=& \sum_{k,q} h_{kq} C^{kq}_J,
\eeqn
where every component $C^{\hat h}_J$ of the resulting vector
having address $J =J(n_1,\cdots,n_k,\cdots,n_q,\cdots,n_M)$, is obtained
as a sum of the $C^{kq}_J$ numbers evaluated for every given $k$ and $q$ according to Eq.~(\ref{one-body_oper}).

Analogously, we group the contribution from all the two-body terms:
\beqn\label{H-psi_two-body}
\hat W \left|\Psi\right> &=&\frac{1}{2}\sum_{k,s,q,l} W_{ksql} \left [b^\dag_k  b^\dag_s b_l b_q\left|\Psi\right> \right ]
= \frac{1}{2} \sum_{k,s,q,l} W_{ksql}  \left|\Psi^{kslq}\right>=\sum^{N_{\mathit{conf}}}_{J}C^{\hat W}_{J}\left|J(\n)\right>, \nonumber  \\
C^{\hat W}_{J}&=&\frac{1}{2}\sum_{k,s,q,l} W_{ksql} C^{kslq}_{J},
\eeqn
where $C^{kslq}_{J}$ are computed using Eq.~(\ref{two-body_oper}) for every given quartet $\{k,s,q,l\}$ of the summation indices.

Finally, the desired result $\hat H | \Psi \rangle$ is obtained as a sum of
the two computed above vectors Eqs.~(\ref{H-psi_one-body}) and (\ref{H-psi_two-body}):
\beqn\label{Hpsi}
\hat H | \Psi \rangle& = &\hat h \left|\Psi\right>+\hat W \left|\Psi\right>=
\sum^{N_{\mathit{conf}}}_{J=1} C^{\hat H}_{J} \left|J(\n)\right>, \\
C^{\hat H}_{J}&=&C^{\hat h}_{J} + C^{\hat W}_{J} \nonumber,
\eeqn
and the respective expectation value of the Hamiltonian is given as a dot product:
\beqn\label{psiHpsi}
\left<\Psi\right| \hat H \left|\Psi\right>\equiv
\left<\Psi\right| \left[ \hat h \left|\Psi\right> \right]+
\left<\Psi\right| \left[ \hat W \left|\Psi\right> \right]
=\sum^{N_{\mathit{conf}}}_{J}C^*_{J} \left(C^{\hat h}_{J}+C^{\hat W}_{J} \right).
\eeqn
These closed form results can be considered as a formal proof that one can operate with state vectors without
construction of the corresponding Hamiltonian matrix. It also justifies the used representation of Eq.~(\ref{MCTDHB_CI}).
Now, instead of the standard matrix representation of the time-(in)dependent Schr\"odinger equations
we can rewrite them in terms of the enumerated configurations, i.e., in a formally ``vectorized'' form:
\beqn\label{NewForm}
i \dot{C_{J}}   & = & C^{\hat H}_{J}, \ \forall  J, \nonumber \\
 C^{\hat H}_{J} & = & E C_{J}, \ \forall  J,
\eeqn
where every $C^{\hat H}_{J}$ element is computed according to Eq.~(\ref{Hpsi}) as a sum of the respective 
one- (\ref{one-body_oper},\ref{H-psi_one-body}) and two-body (\ref{two-body_oper},\ref{H-psi_two-body}) contributions.

Let us conclude.
We implement in the MCTDHB($M$) method the ideas proposed in Ref.~\cite{Mapping_I}
allowing to determine the result of the action of the Hamiltonian written
in second quantized form on a state vector without building up the respective Hamiltonian matrix.
The implemented computational scheme has several advantages comparing to the standard one.
First of all it does not require the (re)evaluation of the Hamiltonian matrix elements in the given many-body basis set
at each time step.
Consequently, there is no need to construct, store and address the Hamiltonian matrix or its elements at all.
Second, the elements of the reduced one- and two-body matrices, needed by the MCTDHB($M$) method
are immediately available as scalar products of the incoming and resulting state vectors,
obtained by acting the respective combination of creation and annihilation operators.
Third, the proposed ideas can be easily extended to three-body and higher-order interactions potentials. 
Last but not least, the proposed ideas are implemented using very effective parallelization schemes,
which are of high demands in modern computational physics.

\section{Many-boson quantum dynamics in triple-well traps}\label{secIII}

Condensation is one of the key physical phenomena taking place in quantum many-boson systems \cite{Pit_Stri_book}.
To characterize a many-boson quantum state one constructs and diagonalizes the reduced one-body density matrix, see Eq.~(\ref{NO}).
According to Penrose and Onsager Ref.~\cite{Cond} the bosonic state is condensed
if only one natural orbital has macroscopic occupation.
On the other hand, when several natural orbitals are macroscopically occupied the system becomes several-fold fragmented \cite{Frag}.
Fragmentation usually takes place in repulsive condensates due to multi-well structure of the external potentials trapping bosonic clouds.
Increasing, for example, the height of the barrier separating two wells
of the double-well trap one can change the ground state property of a weakly-interacting
repulsive many-boson system from almost pure condensed to fully twofold fragmented Refs.~\cite{MCHB,Sipe}.
Analogously, ground state properties of repulsive bosons in optical lattices depend on the intensity of the laser light forming the lattice.
The ground state of weakly interacting bosons in shallow optical lattices is superfluid, i.e., condensed ~\cite{OL1,OL2}.
Increasing the intensity of the optical potential, i.e., its depth, one can drive the system from the superfluid state 
to a completely different, so called Mott insulating state ~\cite{OL1,OL2,OL3}.
Considering an optical lattice as an extreme case of a multi-well trap,
one can associate the fragmented states in multi-well systems with Mott insulating states in optical lattices.
Of course, occupations of the fragments in fragmented states of multi-well systems are macroscopic,
while in optical lattices the Mott insulating states are usually characterized by population of a few bosons per site.

For trap potentials manipulated in time, the quantum dynamics starts to play a very crucial role.
For example, when a repulsive bosonic cloud is split by ramping-up a potential barrier the final state
reveals long-lived oscillatory dynamics between condensed and twofold fragmented states ~\cite{MCTDHB1,MCTDHB_Joerg}.
Similarly, when the intensity of the optical lattice is driven non-adiabatically,
one observes dynamical oscillations between superfluid and Mott insulating states ~\cite{OL3}.
These oscillations are damped.
In the numerical examples of the present section we consider many-boson systems
trapped in a real-space one-dimensional triple-well potential and study their evolutions
by manipulating the shape of the trap, and varying the number of bosons and their interaction strength.
Compared to numerous theoretical and experimental studies of bosonic systems trapped in double-well traps,
studies of bosonic systems in triple-wells are quite scarce.
However, they can provide a deeper insight into the physics of the additional interactions
appearing between neighboring wells due to the presence of the third one.
This makes one step forward to more detailed understandings of the 
dynamical oscillations between superfluid and Mott insulating states taking place in optical lattices.

In this work we use dimensionless units.
To translate the time-dependent many-boson Schr\"odinger equation Eq.~(\ref{MBSE}) to dimensionless units
we divide the Hamiltonian by the energy unit $\frac{\hbar^2}{L^2m}$, where $m$ is the mass of a boson and 
$L$ is a typical length scale of the system.
The dimensionless one-body Hamiltonian then reads $\hat h(x)=-\frac{1}{2}\frac{\partial^2}{\partial x^2} + V(x)$.
The bosons interact via the popular short-range contact repulsive potential
$W(x-x')=\lambda_0\delta(x-x')$ of strength $\lambda_0$, see Refs.~\cite{Pit_Stri_book,Leggett_review,Pethich_book} and references therein.

\subsection{Geometry of triple-well traps and scenario of the dynamics}\label{secIII.1}

Let us consider systems made of $N=12$, 51 and 3003 repulsive bosons confined in a periodic triple-well potential.
We use periodic boundary conditions to ensure that all three wells are equivalent.
For weakly interacting bosons and sufficiently large barrier heights
one expects that the ground state of the system becomes fully threefold fragmented, i.e., 
according to the standard definition \cite{Frag} three natural orbitals in Eq.~(\ref{NO}) acquire macroscopic occupations.
From a perspective of optical lattice physics such a state can be associated 
with a perfect Mott insulating state having occupation of $N/3$ bosons per site.
We use these fully threefold fragmented states in our dynamical studies as initial wave packets for propagation.
At the beginning of the propagation the optical potential has been suddenly shifted and ramped down to some
final value $V_{final}$. We summarize all the trap manipulations as follows:
\beqn\label{Trap}
V(x,t)=V \sin(x/2+\Delta)^2, \, x \in [-3\pi,3\pi) 
\left \{
\begin{matrix}
t<0, \,\,V&=&V_0, \,\, \Delta=0  \\
t\ge 0, \,\,V&=&V_{final}, \,\, \Delta=\pi/4 
\end{matrix}
\right..
\eeqn
Here, $V_0=120$ is the barrier height of the initial trap where fully treefold fragmented states have been prepeared at $t<0$.
$V_{final}=12$ is the barrier height of the final trap where we study the dynamics for $t\ge 0$,
and $\Delta=\pi/4$ is the shift applied to this final potential.

\begin{figure}[ht] 
\includegraphics[width=11cm,angle=-90]{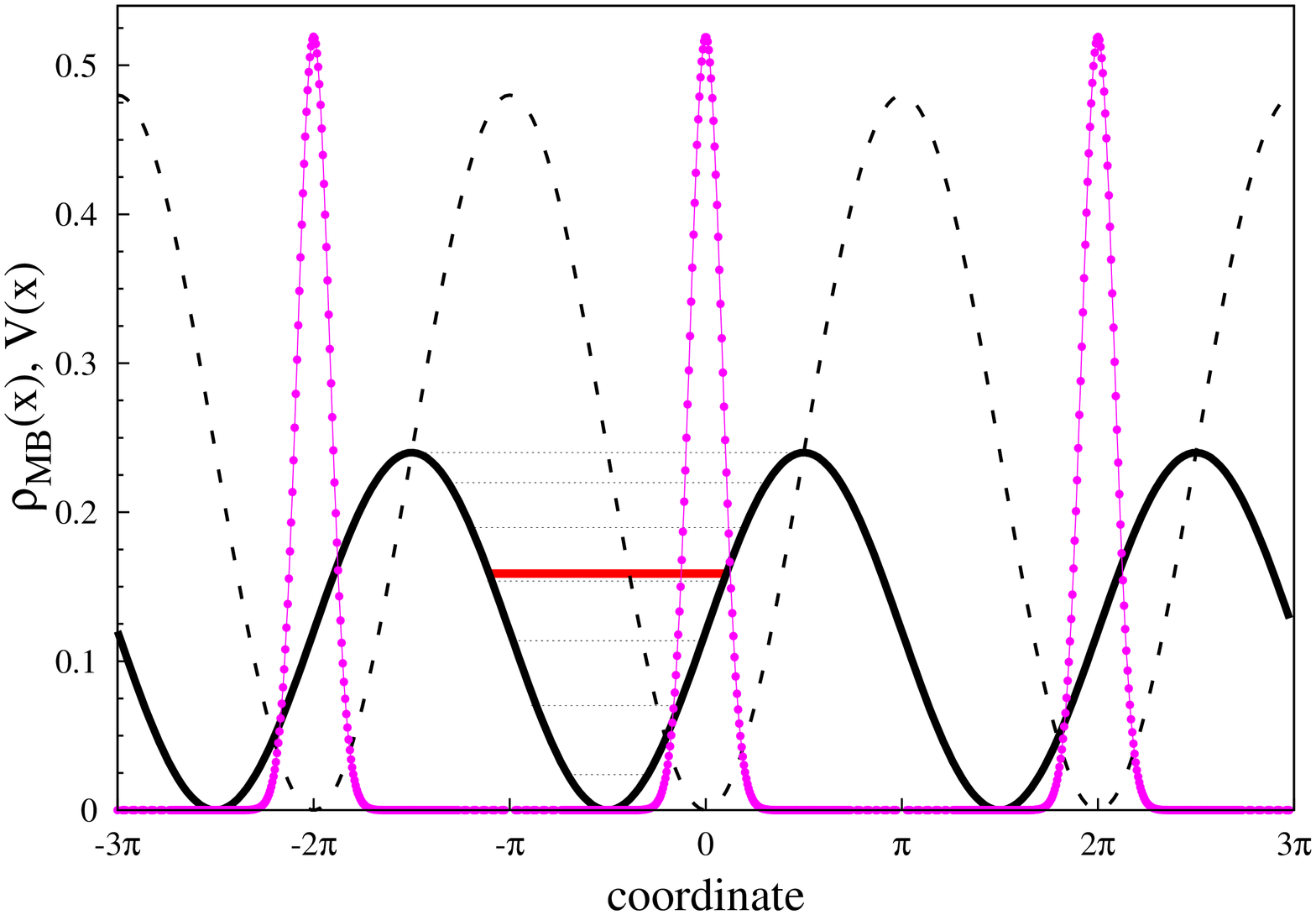}
\vglue 1.0 truecm
\caption [kdv]{(Color online). Scenario of the many-body dynamics in a triple-well periodic trap.
Schematic plots of the initial trap (black dashed line),
normalized initial wave packet density $\rho_{\mathit{MB}}(x)$ (magenta dotted line)
and the final trap potential (solid black line) in which dynamics evolves.
The ground state of $N$ bosons in the initial triple-well periodic potential [see Eq.~(\ref{Trap}) for $t<0$]
is fully threefold fragmented and, therefore, can be viewed as a perfect Mott insulating state with $N/3$ bosons per site.
The final potential obtained at $t=0$ by suddenly shifting and ramping down
the initial trap supports six bands within the wells, plotted as horizontal dashed lines.
In this scenario the initial state is strongly pushed out of equilibrium and
its energy per particle (depicted as a bold red horizontal line) is situated between the fourth and fifth bands.
We consider systems with $N=12$, 51, and 3003 bosons interacting via
repulsive interparticle contact potential of strength $\lambda_0=0.15/(N-1)$.
The quantities shown are dimensionless.
For more details see text.}
\label{Fig.0}
\end{figure}

We propagate the time-dependent Schr\"odinger equation for different numbers $N$ of bosons in real space
within framework of the MCTDHB($M$) method, as described in the previous section \ref{secII}.
The interparticle interaction is chosen to be $\lambda_0=0.15/(N-1)$.
In this case the systems made of different numbers $N$ of bosons become equivalent in the sense that
the fraction of interparticle interaction energy, also associated with the total non-linearity strength $\lambda=\lambda_0(N-1)$,
remains the same for every system. Consequently, the total energies per particle
in these systems are expected to be very close to each other.
That allows us to compare their evolutions.
In Fig.~\ref{Fig.0} we schematically depict the initial and final--shifted trap potentials as well as the initial many-body density.
The topology of the used triple-well trap implies triple degeneracy of the respective one-particle energy levels $\epsilon_i$
of $\hat h(x) \psi(x)=\epsilon_i \psi$, i.e., $\epsilon_0\approx\epsilon_1=\epsilon_2$,
 $\epsilon_3=\epsilon_4\approx\epsilon_5$, etc.
Again, using the analogy between optical lattices and multi-well potentials
we can associate each of these triples of quasi-degenerate levels with a band.
The split between the levels forming a band is smaller for deeper bands.
In Fig.~\ref{Fig.0} we show the lowest-energy levels of each band.
The energy per particle $E_{TD}/N$ of the evolving systems is also depicted by a red solid bold line.
Generally speaking for different $N$ these energies are slightly different, see Table.~\ref{Tab.1},
but at the plotted scale these differences are indistinguishable.
It is important to stress that here we study the dynamics of the interacting bosonic systems in the periodic triple-well potential
where at least six bands are available, see Fig.~\ref{Fig.0}.
The shift $\Delta=\pi/4$ applied to the final potential means that every initial wave packet is pushed far away from equilibrium.
Due to this shift a lot of energy has been pumped into the system, such that
the energy per particle of the initial wave packet is located around the fourth band.
Hence, the system is in a non-equilibrium highly excited state.
After the sudden shift the total energy of the system is conserved because during the wave packet propagation
the final trap potential remains unchanged and the respective Hamiltonian is time-independent and hermitian.

\subsection{Evolution of $N=12$ bosons}\label{secIII.2}

We first study the dynamics of the system made of $N=12$ bosons.
The ground state of such a system becomes fully threefold fragmented already at the barrier height $V_0=7.5$.
However, to be able to relate later the results to those obtained for systems made of different numbers of bosons
we choose as an initial state the ground state at a much higher barrier $V_0=120$.
To get the initial state we propagate the MCTDHB($M$) equations Eqs.~(\ref{MCTDHB_CI},\ref{MCTDHB_PSI})
in imaginary time until the total energy is converged to the desired precision of $10^{-10}$. 
We recall that throughout this work we use dimensionless units.
To simplify the differentiation and integration operations appearing in Eq.~(\ref{MCTDHB_PSI})
we use a discrete variable representation (DVR), see Ref.~\cite{DVR}, for the self-consistent time-adaptive orbitals.
More specifically, each time-dependent orbital is represented in an exponential DVR grid with $N_g=255$ or $N_g=555$ points.
So, in the present computations the linear combination of the $N_g$ primitive exponential functions
define the shape of each orbital at every point in time.
The quality of the used DVR grid can be seen in Fig.~\ref{Fig.0} where the density of the initial state
is plotted as a dotted line: every depicted point corresponds to the actual DVR point in coordinate space ($N_g=255$).
We computed the ground state many-boson wave function at different levels of the MCTDHB($M$) theory,
namely with $M=3$, 6, 9 and 12 self-consistent time-adaptive orbitals.
The obtained ground state energies per particle $E_{GS}/N$ of the initial fully threefold fragmented states are presented
in Table.~\ref{Tab.1}. From this table we clearly see that the ground state energy is converged
with increasing size $N_{\mathit{conf}}$ of the configurational subspace, i.e., with increasing level $M$ of the MCTDHB($M$) theory.

Now, it is worthwhile to clarify the role of self-consistency in the MCTDHB method.
According to Eq.~(\ref{Nconf}), for $N=12$ bosons and three ($M=3$) self-consistent orbitals
the total number of configurations used to describe the many-boson wave function Eq.~(\ref{J_Psi})
is $N_{\mathit{conf}}=91$.
However, since in the numerical computations each orbital is represented as
a sum of $N_g$ exponential DVR functions we can formally expand these sums and compute the total size of the Fock subspace spanned
by these {\it fixed} primitive DVR functions.
For example, for $N_g=255$ exponential functions used in the present computation
the formal dimension of the respective Fock space is huge -- $\binom{N+N_g-1}{N}=203\,656\,421\,123\,930\,640\,320$,
i.e., it is larger than $2\times10^{20}$!
Clearly, this is beyond reach and one cannot carry out even approximately
converged computations using the standard method of fixed orbitals.
The mathematical role of the self-consistent orbitals in MCTDHB
is to contract the fixed primitive DVR basis functions (exponential functions in the present example),
according to the variational principle, and to project, thereby,
the huge configurational subspace to the optimal and compact one.
From a physical point of view this means that every excitation constructed using self-consistent orbitals
actually utilizes a huge number of the primitive excitations built on the fixed DVR functions.
As we see from Table.~\ref{Tab.1} even a modest number of the self-consistent orbitals can provide a very accurate description of the quantum system.

From Table.~\ref{Tab.1} we see that the ground state energy obtained at the three-orbital level of the MCTDHB($M=3$) theory
is very close to those computed using $M=6$, 9, 12 self-consistent orbitals.
We analyze the properties of the respective solutions by constructing and diagonalizing
the reduced one body density [see Eq.~(\ref{NO})] and find that, indeed, only
three natural orbitals are macroscopically occupied: $n_1=n_2=n_3=1/3$.
We conclude that a physically correct description of the ground state is obtained already
at the MCTDHB($M=3$) level.
Further increase of the Hilbert subspace does not lead to significant changes,
but costs, of course, considerable computational efforts (see $N_{\mathit{conf}}$ in Table.~\ref{Tab.1}).
The ground state density obtained for $M=12$ is depicted in Fig.~\ref{Fig.0}.
The ground state density profiles, obtained using $M=3$, 6, 9 self-consistent orbitals
cannot be distinguished from this one and, therefore, are not shown.

\begin{table}
\begin{tabular}{ l|c|c|l|l }
$N$ bosons & $M$ orbitals & Fock space size $N_{\mathit{conf}}$ & $E_{GS}/N$ & $E_{TD}/N$ \\
\hline
$N=12$   &$M=3$  & 91        & 3.8640043 & 7.9216601 \\
         &$M=6$  & 6 188     & 3.8639945 & 7.9216504 \\
         &$M=9$  & 125 970   & 3.8639896 & 7.9216454 \\
         &$M=12$ & 1 352 078 & 3.8639873 & 7.9216431 \\
\hline
$N=51$   & $M=3$ & 1 378     & 3.8679056 & 7.92456997\\
         & $M=6$ & 3 819 816 & 3.8679031 & 7.92456745\\
\hline
$N=3003$ &$M=3$  & 4 513 510 & 3.8689875 & 7.92537683\\
\end{tabular}
\caption{
The energies of the ground and evolving states, and the number of configurations $N_{\mathit{conf}}$
obtained at different levels $M$ of MCTDHB($M$) theory,
i.e., by using different numbers $M$ of self-consistent time-adaptive orbitals.
Results are shown for systems made of $N=12$, 51 and 3003 bosons.
}
\label{Tab.1}
\end{table}

We now turn to the propagation of the initial many-boson wave functions obtained above.
Let us first define the natural time-scale of the problem. 
The energy difference between the second and first bands of the periodic potential $V_{final}(x)$ is 
$\omega=\epsilon_3-\epsilon_0\approx3.52-1.19\approx2.33$, where $\epsilon_0$ and $\epsilon_3$
are the respective lowest-energy levels forming the first and second bands.
This energy difference relates to oscillations of the density in each well (on each site) and
can be used to define the time-scale of the problem $\tau=2\pi/\omega \approx 2.7$. 
We present here the results of the propagation until $t_{total}=400$ which approximately corresponds to $\approx150\tau$
periods of oscillations of the one-site densities.

\begin{figure}[ht] 
\includegraphics[width=10cm,angle=-0]{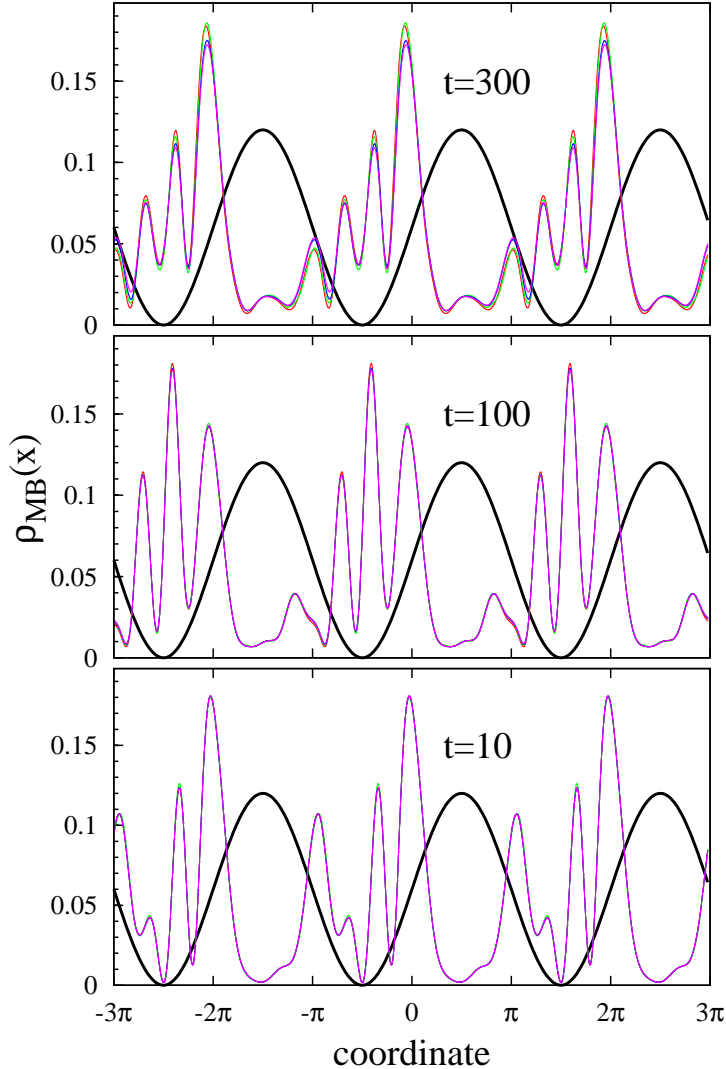}
\vglue 1.0 truecm
\caption [kdv]{(Color online). Time-dependent densities for $N=12$ bosons. 
The time-dependent Schr\"odinger equation is solved for the scenario of Fig.~\ref{Fig.0} for $N=12$ bosons
by using different levels $M$ of MCTDHB($M$) theory. The plotted one-particle densities
$\rho_{\mathit{MB}}(x,t)$ obtained for $M=3,6,9,12$ (red, green, blue and magenta lines) demonstrate 
the convergence of the MCTDHB($M$) results with $M$
for short $t=10\approx3.7\tau$, intermediate $t=100\approx37\tau$ 
and long $t=300\approx111\tau$ propagation times.
Whenever one sees only the magenta line, the curves are all on top of each other.
The energy difference between the second and first bands of the final trap
defines the time-scale of the problem $\tau\approx 2.7$.
The MCTDHB($M=3$) approach utilizing three time-adaptive orbitals provides
an excellent description of the evolving one-body density.
Small discrepancies appearing at longer propagation times can be attributed to quantum depletion,
i.e., excitations to higher orbitals. The quantities shown are dimensionless.}
\label{Fig.I_a}
\end{figure}

Having obtained the initial-state wave function, i.e., respective expansion coefficients and one-particle functions (orbitals),
we use them as initial conditions for integration of the MCTDHB($M$) equations (\ref{MCTDHB_CI}) and (\ref{MCTDHB_PSI})
with $M=3$, 6, 9, 12 orbital sets.
We construct and diagonalize the reduced one-body density Eq.~(\ref{NO}) at each integration time for every set of the orbitals.
The diagonal parts of the reduced one-body densities obtained for a short $t=10\approx3.7\tau$, intermediate $t=100\approx37\tau$ 
and long $t=300\approx111\tau$ propagation times are depicted in the bottom, middle and upper panels of Fig.~\ref{Fig.I_a} respectively.
It is clearly seen that the densities obtained at the MCTDHB($M=3$) level have only marginal deviations
from the $M=6$, 9, 12 ones at all the presented times.
So, the first conclusion is that three time-dependent (self-consistent) orbitals provide the physically correct
description of the density of the evolving many-boson wave packets for quite long propagation times.
The convergence with $M$ is clearly seen in Fig.~\ref{Fig.I_a}.

\begin{figure}[ht] 
\includegraphics[width=11cm,angle=-90]{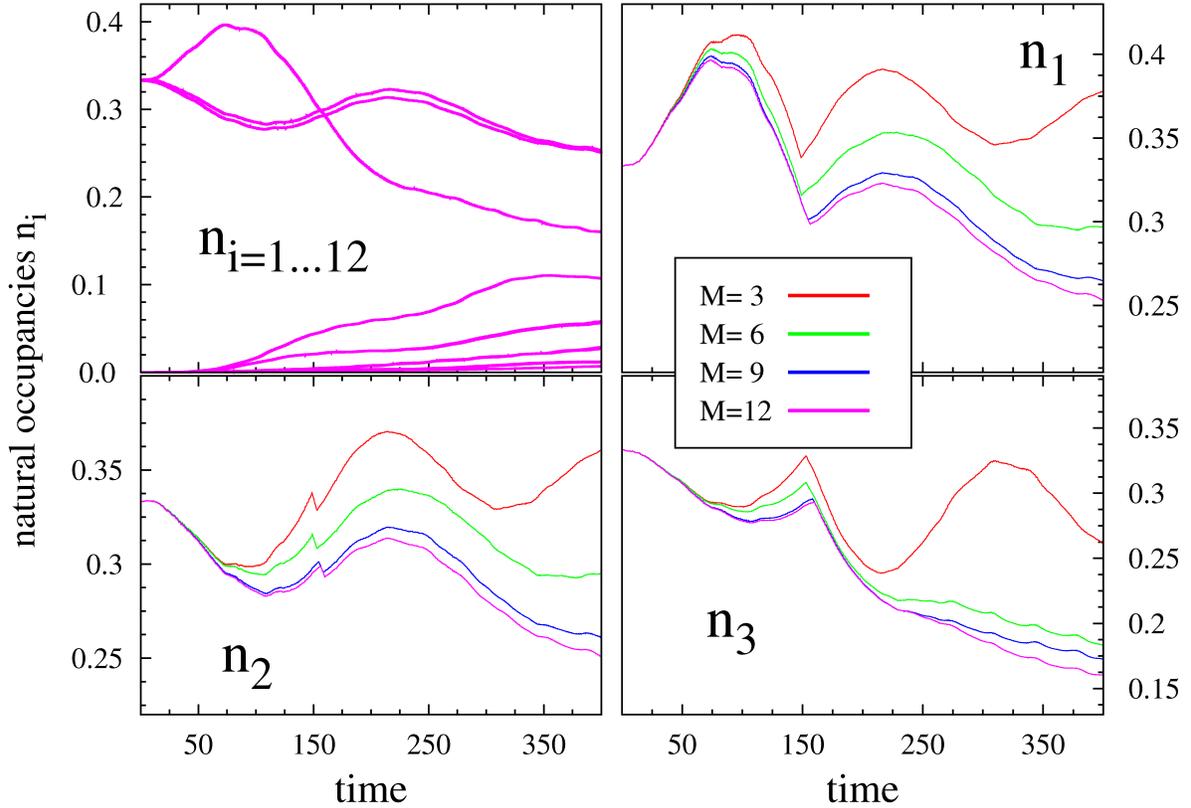}
\vglue 1.0 truecm
\caption [kdv]{(Color online). Properties of the time-dependent many-body wave packets for $N=12$ bosons.
The reduced one-particle density is diagonalized for the system of Fig.~\ref{Fig.I_a}.
The obtained eigenvalues, i.e., natural occupation numbers are plotted as a function of time.
The reference MCTDHB($M=12$) results for evolutions of all twelve natural occupation numbers
are shown in the upper left panel.
The evolutions of the three most-occupied natural orbitals
computed at different levels $M=3,6,9,12$ (red, green, blue and magenta lines)
of the MCTDHB($M$) theory reveal convergence with $M$
as depicted in the upper right, in the lower left and right panels for $n_1$, $n_2$ and $n_3$, respectively.
The long-time oscillations around the pure threefold fragmented state are damped by the quantum depletions,
i.e., excitations to higher orbitals.
Quantum dynamics for the short and intermediate propagation times are
adequately described at the MCTDHB($M=3$) level.
At longer propagation times contributions from higher excited states become more relevant
as time proceeds and more self-consistent orbitals are needed to provide the correct description of these excitations.
Numerical convergence up to about 65 time units is achieved with $M=3$ and up to 150 time units with $M=9$.
The quantities shown are dimensionless.}
\label{Fig.I_b}
\end{figure}

Now, we compare more involved properties of the many-body wave functions. The most important characteristics of
bosonic systems are the degrees of condensation and fragmentation. These quantities correspond
to the values of the natural occupation numbers, computed according to Eq.~(\ref{NO}).
In Fig.~\ref{Fig.I_b} we collect the evolutions of the occupation numbers as functions
of propagation time computed at different levels $M$ of the MCTDHB($M$) theory,
i.e., within different numbers of the time-adaptive orbitals used.
The upper left panel depicts evolutions of the twelve natural occupation numbers obtained at the $M=12$ 
level of computation -- the best converged variational results which we use as a reference.
We recall first that at $t=0$ the initial state is fully threefold fragmented or
using the terminology of optical-lattice physics it is a pure Mott insulating state with $N/3=4$ atoms per site.
Second, the evolution takes place in real space and the shapes of the wave packets are drastically changing in time,
thereby demonstrating the importance of time-dependent orbitals.
Third, the applied shift of the trap potential results in considerable pumping of energy into the system --
compare the energies listed in Table.~\ref{Tab.1} for the initial ground $E_{GS}$ ($t<0$)
and evolving $E_{TD}$ ($t\ge0$) wave packets.
The evolving wave packets are located energetically between the fourth and fifth bands as schematically depicted in Fig.~\ref{Fig.0}.
Nevertheless, despite such a highly non-equilibrium problem, we see that only three natural orbitals
and the respective occupation numbers provide dominant contributions to the physics
of the evolving wave packets, up to rather long propagation times.
In other words, three time-adaptive orbitals are capable to describe the many-boson quantum dynamics of the system.
The evolutions of the three largest occupation numbers computed at different levels $M$ of the MCTDHB($M$) theory,
are compared in the three other panels of Fig.~\ref{Fig.I_b}:
$n_1$ -- in the right upper, $n_2$ -- in the left bottom, and $n_3$ -- in the right bottom panel.
The plotted curves corresponding to the evolutions of $n_1,n_2$ and $n_3$ are not smooth because
the natural occupation numbers obtained by numerical diagonalizing the density matrix Eq.~(\ref{NO}) are sorted according to their values.
Therefore, instead of nicely separated crossing curves we see their connected parts.
In Fig.~\ref{Fig.I_b} one can see that until $t\approx65$ all the MCTDHB($M$) curves for $M=3$, 6, 9, 12 are indistinguishable
on the resolution of the scale shown. In other words, for this time interval which is about $24$ on-site oscillations $\tau$
the three orbital ansatz provides essentially exact description of the many-body dynamics.

Another interesting feature observed from figure Fig.~\ref{Fig.I_b} is that
until $t\approx20\approx7\tau$ the evolving system stays in a quite pure insulating state $n_{i=1,2,3}\approx 1/3$,
despite the fact that the on-site densities undergo drastic changes -- they are split onto three big humps separated
by very deep dips, see lower panel of Fig.~\ref{Fig.I_a}.
With time the many-body dynamics destroys the perfect threefold fragmentation of the initial state.
During the propagation two of the three natural orbitals with largest occupancies
are evolving in time in a very similar manner forming a pair, while the third orbital
is split off and demonstrates a quite different behavior, see upper left panel of Fig.~\ref{Fig.I_b}.
We might associate these evolutions with the structure of the one-particle states of the triple-well trap used.
We recall that each band of the triple-well periodic potential Eq.~(\ref{Trap}) is formed by three discrete eigenstates.
In the lowest bands all of them are almost degenerate, in the higher ones two of the three one-particle
eigenstates are degenerate while the third one is split off.
We thus conclude that the topology of the external trap might be responsible for the observed behavior of the natural occupation numbers
of the evolving many-body wave packets.

The structure of the evolving natural occupation numbers presented in Fig.~\ref{Fig.I_b}
has a very complicated character. We can consider two main scales:
the general leading trend -- gradual long-time oscillation with a period of $t\approx270\approx 100\tau$,
and small modulating short-time oscillations on top of it.
Comparing, first, these small modulating oscillations for different $M$ we conclude that they are
nicely reproduced at all the presented levels of the MCTDHB($M$) computations, starting already at $M=3$.
The long-scale dynamics might be viewed as a damped oscillations around perfect threefold fragmented state.
These oscillations may be attributed to periodic attempts to develop coherence in the system.
Alternatively, using optical-lattice terminology one may say that this behavior mimics the oscillations
between Mott insulating and superfluid states.
Till $t\approx65$ all the results practically coincide -- one can not see the deviations at the presented scales of Fig.~\ref{Fig.I_b}.
For $t>65$ the discrepancy in the long time behavior between $M=3$ and $M=6$, 9, 12 computations starts to develop,
remaining nevertheless quite modest. This, again, indicates that the essential part of the evolution 
is nicely described at the MCTDHB($M=3$) level.

Let us now analyze the physical origin of the observed imperfection (discrepancy) 
in the long time behavior between the $M=3$ and $M=6$, 9, 12 results for the evolving natural occupation numbers depicted in Fig.~\ref{Fig.I_b}.
We recall first that in all the presented computations the total energy of the evolving systems is preserved
and, second, temperature does not enter the computations at all.
The key difference between $M=3$ and $M>3$ MCTDHB($M$) computations is the number of time-adaptive orbitals available for dynamics,
i.e., the number of excited states taken into account (see respective dimensions $N_{\mathit{conf}}$ listed in Table.~\ref{Tab.1}).
We can suppose that excitations to higher orbitals, not available at the MCTDHB($M=3$) level, are responsible for the discrepancy.
On the other hand, these quantum excitations deplete the occupations of the three leading fragments and, hence,
we can call this process quantum depletion.
The proposed explanation can be directly verified by comparing the details of the three-orbital
and $M>3$ computations. In the $M>3$ cases the depletion is naturally described by populating higher time-adaptive orbitals.
Indeed, the MCTDHB($M$) results using $M=6$, 9 and 12 time-adaptive orbitals are very close to each other and
show that the populations of the higher $M>3$ natural orbitals increase with time.
This can be seen from the left upper panel of Fig.~\ref{Fig.I_b} where all the natural occupation numbers for $M=12$ case are plotted.

Now, we elaborate more on the role of depletions in the three-orbital MCTDHB($M=3$) description.
We recall that every time-adaptive orbital is a sum of primitive DVR functions (plane waves in this study).
Hence, an excitation involving only a few time-dependent optimized orbitals actually contracts a
huge number of excitations between the primitive plane waves.
So, due to this mutual self-consistency of the MCTDHB method large fraction of quantum depletion
is simulated by readjustments of the orbitals' shapes already at the three-orbitals level.
Since the total number of particles and self-consistent orbitals are kept fixed,
the ``depleted'' populations can only be redistributed between available self-consistent orbitals,
which is reflected as changes of the natural occupation numbers.
Of course, at longer propagation times contributions from higher excited states become more and more relevant
and a larger number of the time-adaptive orbitals are needed to provide a more accurate description of these excitations.
This explains the small differences in the densities between $M=3$ and $M>3$
cases seen in Fig.~\ref{Fig.I_a} at long propagation times ($t=300$).

In the scenario studied here the processes of quantum depletion play a very important role -- they
lead to a considerable damping of the long-scale oscillation around the perfect threefold fragmented state.
The general conclusion is that at the MCTDHB($M=3$) level of description almost all key physical features of the evolution 
are accurately reproduced, while numerically converged results are obtained using more time-adaptive orbitals.

\subsection{Evolution of $N=51$ bosons}\label{secIII.3}

Now we discuss the second numerical example with $N=51$ repulsive bosons
trapped in the same triple-well potential as before, Eq.~(\ref{Trap}).
As an initial state we take, again, the ground state of the system at $V_0=120$ and $\lambda_0=0.15/(N-1)$. Similarly to the $N=12$ case,
the ground state becomes fully treefold fragmented $n_{i=1,2,3}=1/3$ at smaller barrier heights starting at $V_0=10.0$.
We propagate the MCTDHB($M$) equations (\ref{MCTDHB_CI}) and (\ref{MCTDHB_PSI}) using $M=3$ and $M=6$ time-adaptive orbitals
in imaginary time to get the ground state expansion coefficients and respective one-particle functions.
The obtained ground state energies per particle are listed in Table.~\ref{Tab.1}.
First of all, we see a convergence of the ground state energy with $M$.
Second, comparing the results for $N=12$ and $N=51$ we conclude that these systems
are indeed very similar from the point of view of energy per particle.

\begin{figure}[ht] 
\includegraphics[width=11cm,angle=-0]{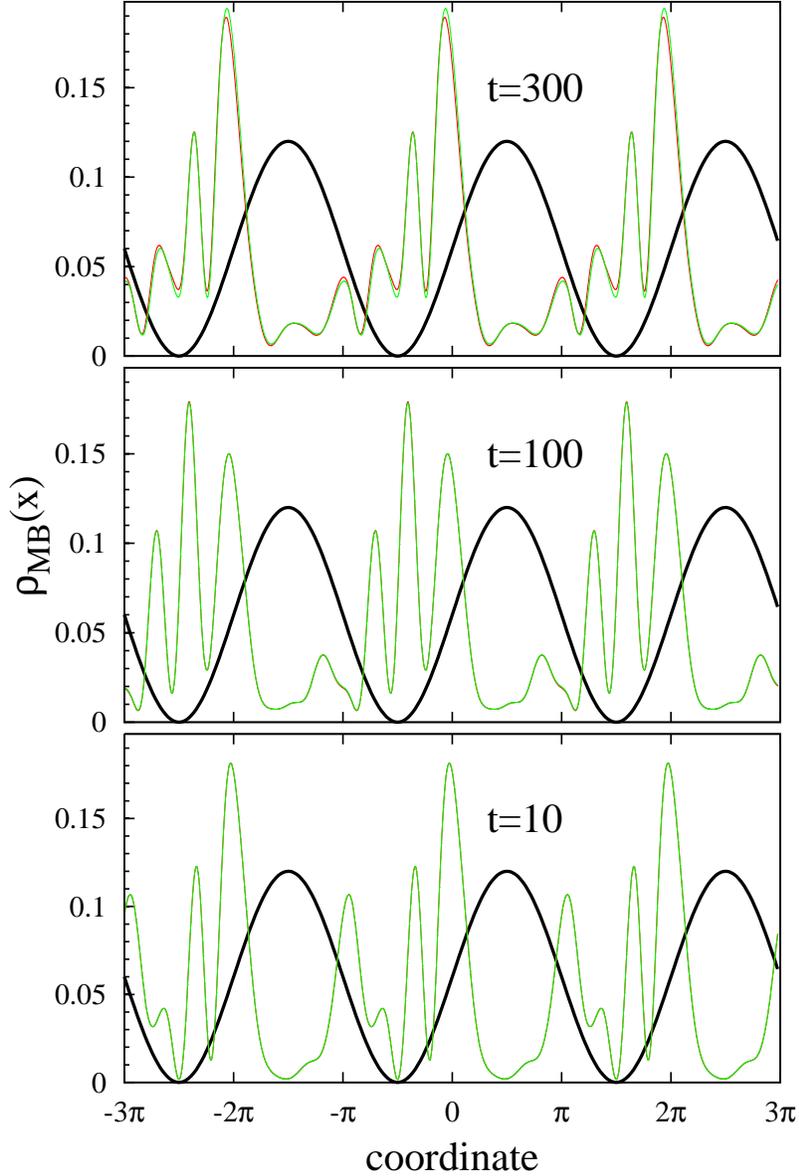}
\vglue 1.0 truecm
\caption [kdv]{(Color online). Many-body dynamics of $N=51$ bosons and the scenario of Fig.~\ref{Fig.0}. Shown are the
one-particle densities $\rho_{\mathit{MB}}(x,t)$ computed at the $M=3$ and $M=6$ levels (red and green lines) of the MCTDHB($M$) theory
for short $t=10\approx3.7\tau$, intermediate $t=100\approx37\tau$ and long $t=300\approx111\tau$ propagation times.
Whenever one sees only the green line, the curves are on top of each other.
Very small distinctions between three- and six-orbitals' computations start to appear at long propagation times ($t=300$).
The quantities shown are dimensionless.}
\label{Fig.II_a}
\end{figure}

\begin{figure}[ht] 
\includegraphics[width=11cm,angle=-90]{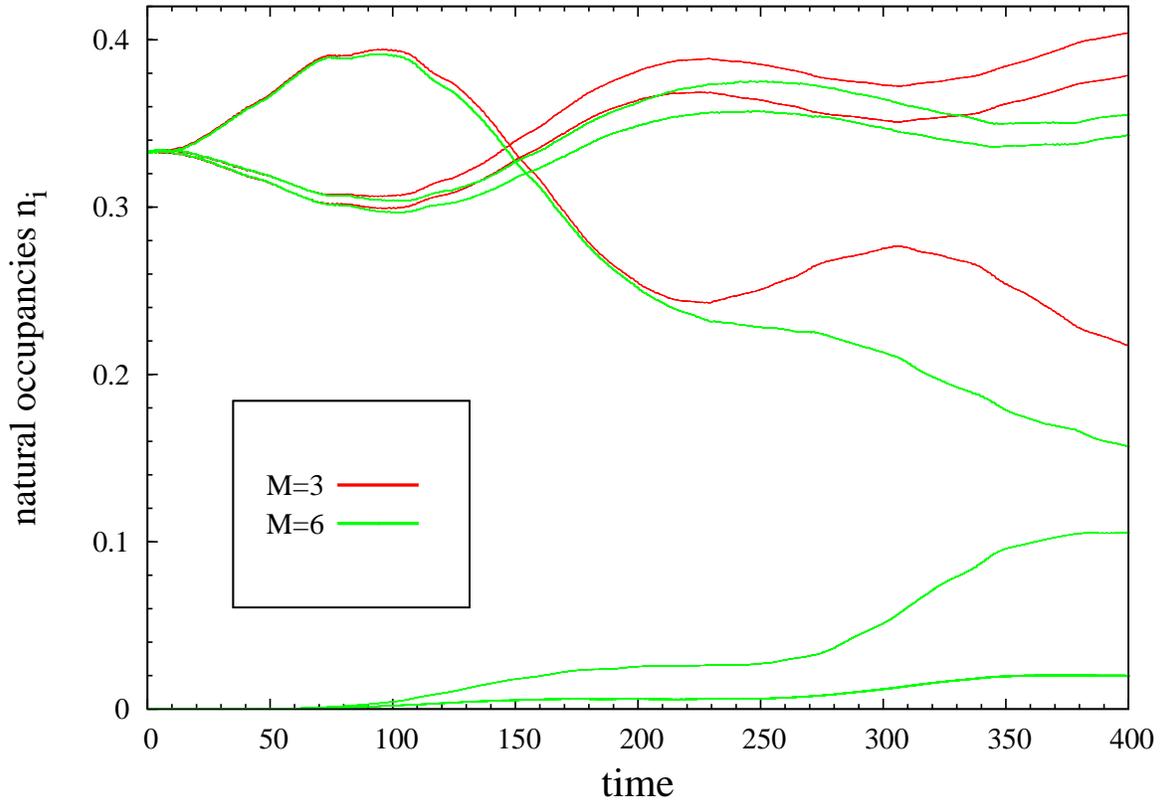}
\vglue 1.0 truecm
\caption [kdv]{(Color online). Properties of the evolving many-body wave packet for $N=51$ bosons.
Evolutions of the natural occupation numbers $n_i$ obtained by using three- and six-orbital (red and green lines) levels of the 
MCTDHB($M$) theory are plotted as a function of time.
For the short and intermediate propagation times the three-orbital ansatz is quantitative,
at longer propagation times quantum depletions become more and more relevant,
and more time-adaptive orbitals are needed.
Note that the one-particle densities computed with $M=3$ and $M=6$ do not defer
from each other even for the longer propagation time (see Fig.~\ref{Fig.II_a}). The quantities shown are dimensionless.}
\label{Fig.II_b}
\end{figure}

The next step is dynamics -- we ramped down the periodic potential suddenly to the depth $V_{final}=12$ 
of the triple-well and shift it to $\Delta=\pi/4$, see Eq.~(\ref{Trap}), and then perform the real-time propagation.
We again follow the evolution of the many-boson wave function and construct and diagonalize the reduced one-body density matrix 
according to Eq.~(\ref{NO}) at each point in time.
The representative densities obtained at $t=10\approx3.7\tau$, $t=100\approx37\tau$ and $t=300\approx111\tau$, i.e., at
the short, intermediate and long evolution times are depicted in Fig.~\ref{Fig.II_a}.
The densities computed at the $M=3$ and $M=6$ levels of the MCTDHB($M$) theory are very similar at all the propagation times,
indicating that three time-adaptive MCTDHB($M=3$) orbitals provide an adequate
description of the evolution in this case as well, as far as the density is considered.
The one-particle densities obtained at the $M=3$ and $M=6$ levels are plotted in Fig.~\ref{Fig.II_a} using red and green colors.
In Fig.~\ref{Fig.II_b} we plot the evolutions of the corresponding natural occupation numbers as functions of time
using the same color scheme -- red and green curves represent MCTDHB($M$) results for the $M=3$ and $M=6$ computations.

The general behavior of the system with $N=51$ bosons is quite similar to that discussed above for $N=12$.
Until $t\approx20$ the system continues to stay in a quite pure threefold fragmented state.
During this time the density, however, drastically changes, see the density profile at $t=10$
depicted in the lower panel of Fig.~\ref{Fig.II_a}.
Afterwards, the occupations of two of the three initially equally occupied natural orbitals are reduced.
The largest occupation number reveals an opposite tendency -- it borrows particles from
the other two natural orbitals, increasing thereby its own natural occupation.
This picture described at the $M=3$ level is confirmed by the computation with $M=6$ orbitals.
Thus, short and intermediate time evolutions ($t \lesssim 75$) are identically
described at the $M=3$ and $M=6$ levels of MCTDHB theory.
Quantum depletions -- excitations from three macroscopically-populated self-consistent time-adaptive orbitals to higher ones --
also become relevant for this system as time proceeds and appear as a gradual decrease of the occupations of all the three fragments.
As we have already mentioned above, the considered systems of $N=12$ and $N=51$ bosons are similar
because we took the same trap potentials, the ``same'' initial state and very similar interparticle energies defined by the same $\lambda$.
We can compare the behavior of the depletions as a function of time for these systems by
comparing the occupations of higher $M>3$ natural orbitals.
These are presented in the left upper panel of Fig.~\ref{Fig.I_b} for $N=12$
and in Fig.~\ref{Fig.II_b} for $N=51$. We conclude that quantum depletion develops slower for the system made of a larger number of bosons.

\subsection{Evolution of $N=3003$ bosons}\label{secIII.4}

Let us now discuss the last numerical example, where we consider dynamics of $N=3003$ repulsive bosons with 
contact interaction strength $\lambda_0=0.15/(N-1)$ trapped in the same triple-well potential
as before. To obtain a perfect threefold fragmented, i.e., Mott insulating state with 1001 bosons per site,
the optical depth of the periodic potentials has to be very large, $V_0=120$.
We first propagate the MCTDHB($M$) equations with $M=3$ time-adaptive orbitals
in imaginary time to get the ground state solution of this system. Using Eq.~(\ref{NO}) we verify that
the obtained state is indeed threefold fragmented --  the respective natural occupation numbers are $n_{i=1,2,3}=1/3$.
Then, we suddenly shift the potential to $\Delta=\pi/4$ and reduce its amplitude to $V_{final}=12$ as before, see Eq.~(\ref{Trap}),
and apply the real-time propagation scenario. The computed energies of the initial system $E_{GS}$
as well as of the evolving wave packet $E_{TD}$ are presented in Table.~\ref{Tab.1}.

\begin{figure}[ht] 
\includegraphics[width=11cm,angle=-0]{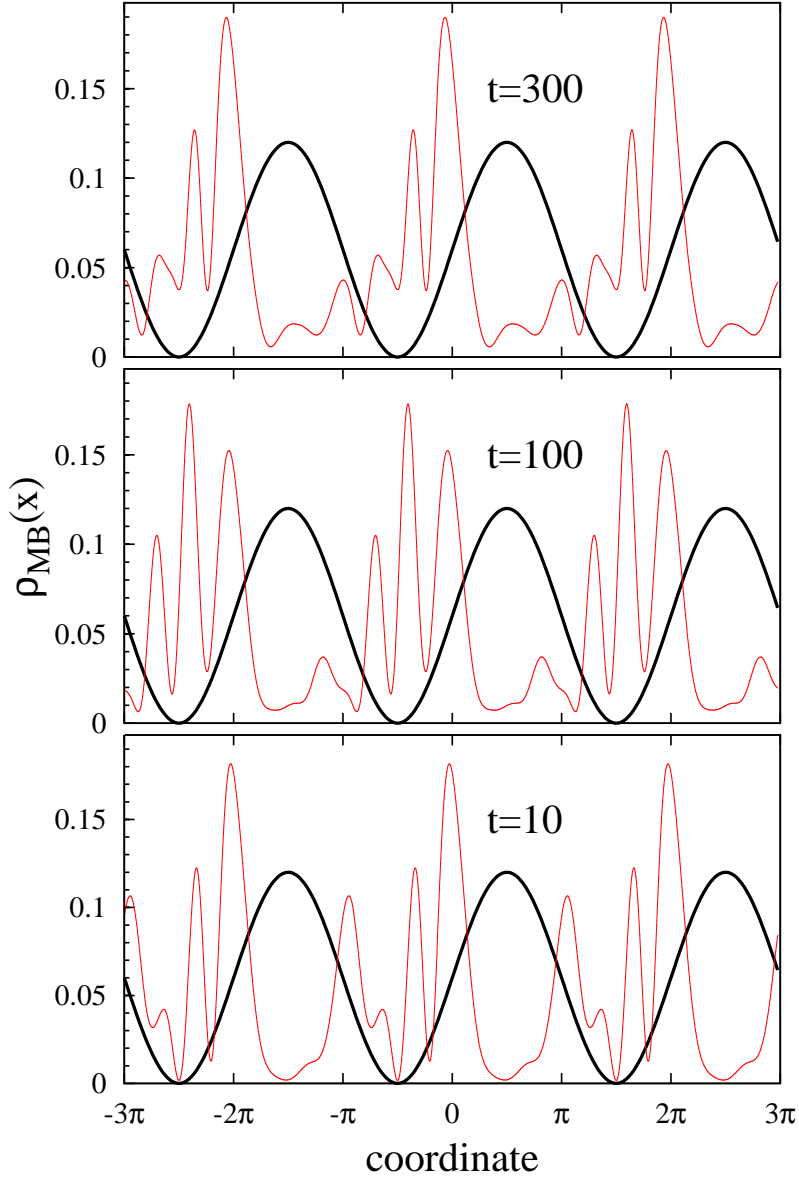}
\vglue 1.0 truecm
\caption [kdv]{(Color online). Many-body dynamics of $N=3003$ and the scenario of Fig.~\ref{Fig.0}.
Shown are the MCTDHB($M=3$) one-particle densities $\rho_{\mathit{MB}}(x,t)$ obtained for the short
$t=10\approx3.7\tau$, intermediate $t=100\approx37\tau$ and long $t=300\approx111\tau$ propagation times.
In this work we keep the interparticle interaction energy, i.e., non-linearity parameter $\lambda=\lambda_0(N-1)$
the same for all the considered systems and choose $\lambda_0=0.15/(N-1)$.
The time-dependent densities presented in this figure for $N=3003$ are quite similar to these
depicted in Figs.~\ref{Fig.I_a} and \ref{Fig.II_a} for $N=12$ and $N=51$ bosons, respectively.
The systems made of different number of particles but characterized by the same $\lambda$
reveal quite similar density dynamics. The quantities shown are dimensionless.}
\label{Fig.III_a}
\end{figure}

\begin{figure}[ht] 
\includegraphics[width=11cm,angle=-90]{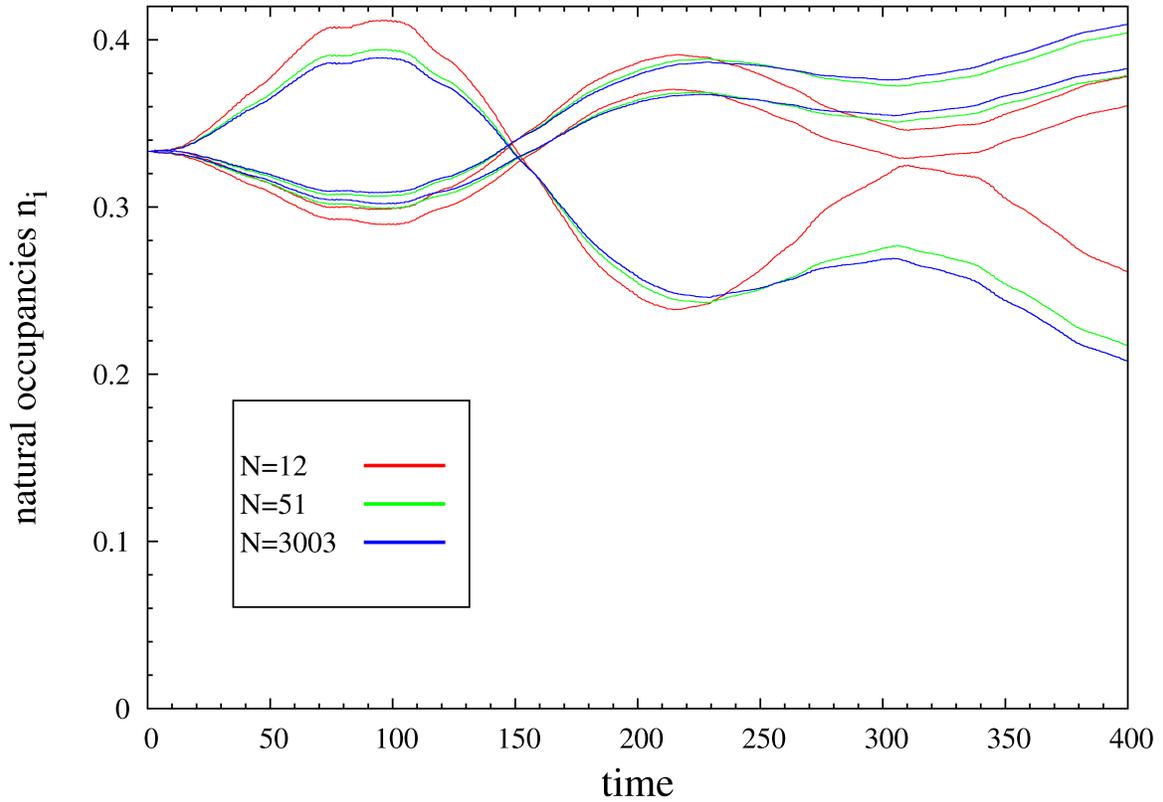}
\vglue 1.0 truecm
\caption [kdv]{(Color online).
Properties of time-dependent many-body wave packets for $N=3003$ bosons.
Evolutions of the three natural occupation numbers $n_{i=1,2,3}$ obtained within the MCTDHB($M=3$) theory
are plotted as a function of time. For comparison also the results for $N=12$, and 51 bosons are shown.
The evolution of the systems made of $N=51$ and $N=3003$ bosons are much closer to each other
than to the $N=12$ ones. The many-body properties seem to be saturated to some average values already at $N=51$.
The quantities shown are dimensionless.}
\label{Fig.III_b}
\end{figure}

Now, we analyze the results of the real-time propagation.
In Fig.~\ref{Fig.III_a} we plot the densities at the short $t=10\approx3.7\tau$, intermediate
$t=100\approx37\tau$ and long $t=300\approx111\tau$ propagation times.
Comparing these results with the respective densities depicted in Fig.~\ref{Fig.I_a} and Fig.~\ref{Fig.II_a}
for the bosonic systems made of $N=12$ and $N=51$ particles we
conclude that all three systems exhibit very similar behaviors.
We apply Eq.~(\ref{NO}) to analyze the properties of the propagating wave packet and depict in Fig.~\ref{Fig.III_b}
the obtained evolutions of the three available natural occupation numbers using blue lines.
Two of the three, initially equally occupied natural orbitals start to donate the bosons, i.e., decrease their occupancies,
while the third one accepts the particles, increasing its own occupation.
Then, around $t\approx 100$, the tendency changes and the most-occupied
natural orbital starts to donate bosons, transferring them back to the two other orbitals.
A signature of this imperfect oscillatory behavior can be observed
during all the computed propagation times depicted in Fig.~\ref{Fig.III_b}.
For comparison in this figure we also plot the natural occupation numbers obtained at the $M=3$ level of the MCTDHB($M$) theory
for the systems made of $N=12$ and $N=51$ bosons with red and green colors, respectively.
This allows us to analyze the $N$-dependence of the evolutions.
The key observation is that evolutions of all the three systems are qualitatively the same.
Moreover, the evolutions of the systems made of $N=51$ and $N=3003$ bosons are much closer to each other
than to the $N=12$ ones. Therefore, we might speculate that for the present studies the many-body properties are
saturated to some average values already at $N=51$. Hence one can expect that key aspects of the dynamics of the initially threefold
fragmented systems with larger $N$ will be very similar to those obtained for $N=51$.

Let us now summarize all three studies. We have considered systems made of $N=12$, $N=51$ and $N=3003$ particles
choosing the strength of the contact interparticle interaction potential 
such that the non-linearity parameter $\lambda=\lambda_0(N-1)$ is the same for all systems.
We have solved the many-particle time-dependent Schr\"odinger equation
using the scenario of Eq.~(\ref{Trap}) (see also Fig.~\ref{Fig.0}) for all the considered systems.
Comparing the diagonal parts of the reduced one-particle density matrices, i.e., the densities computed
at different propagation times and depicted for $N=12$, $N=51$, and  $N=3003$ bosons
in Figs.~\ref{Fig.I_a}, \ref{Fig.II_a} and \ref{Fig.III_a}, respectively,
we observe that the presented densities look very similar at the short, intermediate and long propagation times.
We also verify that the energy per particle of the considered systems are very close to each other,
as one can see by comparing the data collected in Table.~\ref{Tab.1}.
Hence, we can conclude that the non-linearity parameter $\lambda=\lambda_0(N-1)$ is useful to characterize the global
behavior of the densities in the systems made of different numbers of particles,
at least for the dynamical studies of the presented type.
The compared results for the time-dependent densities are quite similar, but more involved properties reveal differences.
Indeed, the comparison of the properties of the evolving wave packets, e.g., evolutions of the natural occupation numbers
presented in Figs.~\ref{Fig.I_b}, \ref{Fig.II_b} and \ref{Fig.III_b} for $N=12$, $N=51$, and $N=3003$, respectively,
shows that global tendencies are $\lambda$ dependent, while the depletions, i.e., excitations from
the three most-occupied orbitals to higher ones depend in addition
on total number $N$ of the bosons in the system as well
as on their interaction strength $\lambda_0$.
We conclude that in the presented study the dependence of the properties on the number of particles $N$
explicitly appears at longer propagation times and is related to the quantum depletion which is also found to grow with time.

\section{Summary and conclusions}\label{secIV}

Recently, in Ref.~\cite{Mapping_I} we proposed
a novel, effective and general theory to construct the result of action of any
many-particle operator represented in the second quantized form on a state vector,
without resorting to matrix representation of operators and even to its elements.
In the present work we adopt the bosonic version of this theory to the MCTDHB method.
We first discuss the practical realization of the proposed ideas in the MCTDHB($M$) computational scheme
and address its parallel implementation in some details.
Using this new tool we study highly non-equilibrium dynamics of repulsive systems made of $N=12$, 51, 3003 particles
in real-space triple-well traps. As an initial wave packet we take the fully threefold fragmented quantum state,
strongly displace it out of equilibrium and then, suddenly, reduce the depth of the potential wells.
From the perspective of optical-lattice physics this scenario can be seen as a propagation of the initially pure Mott insulating state,
strongly displaced out of the potential minima.
As a result of such a sequence of trap manipulations at least four bands
of the periodic trap potential become energetically open, i.e., available for excitations.
We have shown that quantum dynamics for this scenario is very complicated.
First, we have demonstrated the convergence of the MCTDHB($M$) approach with $M=3$, 6, 9, 12 self-consistent time-adaptive orbitals
by applying the described above scenario to study long-time dynamics of $N=12$ bosons.
Next, we see that a quite accurate description of the evolution for short and even intermediate propagation times
is obtained already at the MCTDHB($M=3$) level of theory.
This result provides, thereby, a confirmation of an intuitive supposition that many-body dynamics in three-well traps can qualitatively
and sometimes quantitatively be described by using three self-consistent time-adaptive orbitals.
We stress here the last point -- each used self-consistent time-adaptive orbital actually contracts
very many fixed (primitive) functions ($N_g=255$ or 555 plane waves in our work), and, therefore, the MCTDHB many-body wave function
constructed by using these few self-consistent time-adaptive orbitals
spans a huge effective Fock space.

As a next step we compute the dynamics for the systems made of $N=51$ and $N=3003$ bosons
and characterized by the same total non-linearity parameter $\lambda\equiv\lambda_0(N-1)$.
In the region of the trap/state manipulations considered here,
the total non-linearity $\lambda$ appears to be quite a useful parameter.
Because the systems made of different numbers of atoms but with the same $\lambda$ 
demonstrate very similar dynamical behavior at least for the evolving density profiles
of the initially three-fold fragmented (Mott insulating) states.
More involved properties, like natural occupation numbers are more sensitive 
to the individual $\lambda_0$ and $N$ parameters, especially for a small number of particles and at longer propagation times.
The evolutions of the natural occupation numbers show up a tendency to a long-time oscillatory
behavior around the initially perfect threefold fragmented state.
However, quantum depletions, i.e., excitations to higher orbitals lead to suppression or damping of these oscillations.
For the short and intermediate propagation times this process is adequately reproduced already at the MCTDHB($M=3$)
level due to mutual self-consistency of the time-adaptive orbitals used.
However, for longer propagation times more self-consistent orbitals are needed for the correct description
of the depletions and many-body dynamics. At higher levels of the MCTDHB($M>3$) description, e.g., within $M=6$ self-consistent orbitals,
the depletions are naturally included and with $M=9$, 12 orbitals they are accurately described.

The MCTDHB theory augmented by the proposed, developed and implemented ideas for
effective operations in the bosonic Fock space allows for a faster solution
of the time-dependent many-boson Schr\"odinger equation in larger configurational spaces and for longer propagation times.
This opens new perspectives for quantitative descriptions of the quantum many-body
dynamics of larger systems, as recently announced through the 
numerically exact studies of the one-dimensional bosonic Josephson junction problem \cite{BJJ}.

\begin{acknowledgments}
Financial support by the Deutsche Forschungsgemeinschaft is gratefully acknowledged.
\end{acknowledgments}

\end{document}